# Atom probe tomography of hydrated biomacromolecules: preliminary results

Shuo Zhang[1], Leonardo Shoji Aota[1], Mahander P. Singh[1], Eric V. Woods[1], Fantine Périer Jouet[1], Tim M. Schwarz[1], Baptiste Gault[1,2,*].

[1] Max-Planck Institute for Sustainable Materials (formerly Eisenforschung), Max-Planck-Strasse 1, 40237 Düsseldorf, Germany

[2] Department of Materials, Royal School of Mines, Imperial College

[*] Corresponding author: b.gault@mpie.de

## Abstract

The folding and structure of biomacromolecules depend on the 3D distributions of their constituents, which ultimately controls their functionalities and interactions with other biomacromolecules. Atom probe tomography (APT) with its unparalleled compositional sensitivity at nanoscale spatial resolution, could provide complementary information to cryo-electron microscopy, yet routine APT analysis of biomacromolecules in their native state remains challenging. Here, a ferritin solution was used as a model system. Following plunge freezing in liquid nitrogen, cryogenic lift-out and cryo-APT analysis were performed. Elements from the ferritin core and shell are detected yet particles seem destroyed. We hence demonstrate the feasibility of preparing and analyzing bulk hydrated biological samples using APT, however, the cooling was too slow to vitrify the solution. This caused irrecoverable damage to the protein shell surrounding the ferritin particles due to ice crystal formation. We report on preliminary data from high-pressure frozen (HPF) deionized (DI) water, demonstrating a proof-of-principle experiments that intact biomacromolecules could be analyzed through a similar workflow in the future. We report on many trials (and errors) on the use of different materials for substrates and different substrate geometries, and provide a perspective on the challenges we faced to facilitate future studies across the community.

## 1 Introduction

Understanding biological macromolecules' 3D structures and compositions is crucial for unravelling their functions, since structure is directly related to their biophysical and biochemical properties. A better understanding of the interactions between macromolecules and/or host tissues will provide opportunities to design novel biochemical compounds and processes (Adrian et al., 1984; Bodakuntla et al., 2023). Scanning electron microscopy (SEM) (Dumoux et al., 2023) and transmission electron microscopy (TEM) (Frank, 2016; Dubochet, 2016) have been used extensively to characterize the structure biological materials with resolutions from sub-micron to near-atomic, in part facilitated by progress in cryogenically-enabled focused-ion beam (cryo-FIB) microscopes (Hayles et al., 2007; de Winter et al., 2020). Atom probe tomography (APT) could be a promising and effective approach to investigate the microstructure and chemical composition of biological materials, with a sub-nanometer resolution, and can complement TEM and SEM to gain new insights in the nanoscale composition and possibly the structure of biomacromolecules.

APT requires the specimen to be shaped as a sharp needle with a tip diameter below 150 nm, to produce a high electrostatic field at the specimen's apex in the range of $10^{10}$ V.m$^{-1}$, using moderate high voltages in the range of 2 – 12 kV (Lefebvre-Ulrikson et al., 2016). Such fields enable the removal of atoms from the specimen's surface in the form of ions, a process termed field evaporation. The ions then fly toward a position-sensitive detector, which collects each ion's impact position and time of flight (Müller et al., 1968; Blavette et al., 1993). This information is used to build a 3D reconstruction that facilitate chemical composition mapping of the original specimen (Gault et al., 2021). In previous studies, APT has primarily been used for the analysis of metallic materials and semiconductors, but recent studies have been reported for dehydrated biological materials, including amyloid (Rusitzka et al., 2018), ferritin (Perea et al., 2016; Greene et al., 2010), and embedded proteins (Adineh et al., 2017; Sundell et al., 2019). However, the dehydration or embedding process can damage and distort the biological structure of interest and may not represent their pristine structure.

Recently, APT of hydrated specimens such as pure water (Schwarz et al., 2020), ferritin (Qiu et al., 2020), amyloid fibrils (Woods et al. 2024) and various solutions (El-Zoka et al., 2020; Schwarz et al., 2021, 2022; Eric V Woods et al., 2023; E. V. Woods et al., 2023) have been reported. However, the specimen transfer, data collection and analysis remain extremely challenging, especially for frozen-hydrated specimens. The development of cryo-transfer devices and gloveboxes (Stephenson et al., 2018; McCarroll et al., 2020; Perea et al., 2017; Gerstl et al., 2017), which allowed transfer specimens under cryogenic und UHV conditions between different instruments enabled the progress towards hydrated samples. The advantage of using the APT to characterize hydrated samples is that it can in principle maintain the near-native state, i.e. minimize damage to the structure of biological molecules. However, one of the main difficulties in analyzing hydrated samples is the rapid freezing of the samples directly into the amorphous state to avoid any damage and structural changes to the biological structure. Progress in specimen preparation was made by using the re-deposition cryo-lift-out (LO) method, which enables the preparation of multiple specimen on a single support, typically a commercial Si coupon, even from bulk liquid samples at cryogenic temperature (Douglas et al., 2023; Eric V Woods et al., 2023). In this newly developed protocol, instead of using *in-situ* Pt-deposition by using a gas-injection system (GIS) to enhance the mechanical stability of the junction, metallic Cr is sputtered from a target brought to the vicinity of the junction between the material of interest and the substrate, and was re-deposited to stabilize the junction for further analysis using APT (Eric V Woods et al., 2023).

Here, we aimed to use cryo-APT to study hydrated ferritin buffer solution, using the specimen preparation approach outlined in (Eric V Woods et al., 2023). A solution containing ferritin particles in water was frozen by manually plunge freezing in liquid nitrogen (LN2) as described in previous work from our group (El-Zoka et al., 2020; Schwarz et al., 2024). Afterwards, specimens were prepared by using cryo-LO on a cryo-(Ga)FIB, followed by cryo-transfer for APT analysis through a commercial cryo-vacuum carry shuttle known as a suitcase (Ferrovac GmbH). The suitcase itself is a portable ultrahigh vacuum ($<10^{-7}$ mbar) chamber in which the sample holder sits on a plate connected to a liquid nitrogen dewar and hence maintained at cryogenic temperature (below approx. - 180 °C). Multiple metal substrates were tried for plunge freezing. Specimens only from the matrix solution, or the solution and the substrate were prepared and analyzed. During APT measurement, parameters including the laser pulsing rate and laser pulse energy, were varied to optimize the data quality. Mass spectra and 3D elemental maps reveal the iron distributions, and the interface between the liquid and the metal substrate. The results show that sodium is distributed throughout the specimens, with a higher concentration at the metal/liquid interface. We demonstrate the feasibility of preparing and analyzing bulk hydrated biological samples using APT, however, plunge freezing in LN$_2$ is too slow to vitrify the solution into its vitreous state. This caused irrecoverable damage to the protein shell surrounding the ferritin particles due to the ice crystal formation. Finally, we demonstrate that the same experimental workflow can be used to analyze pure ice water (deionized water) prepared by high-pressure freezing (HPF), paving the way for probing biological specimens in their near-native state.

## 2 Materials and Methods

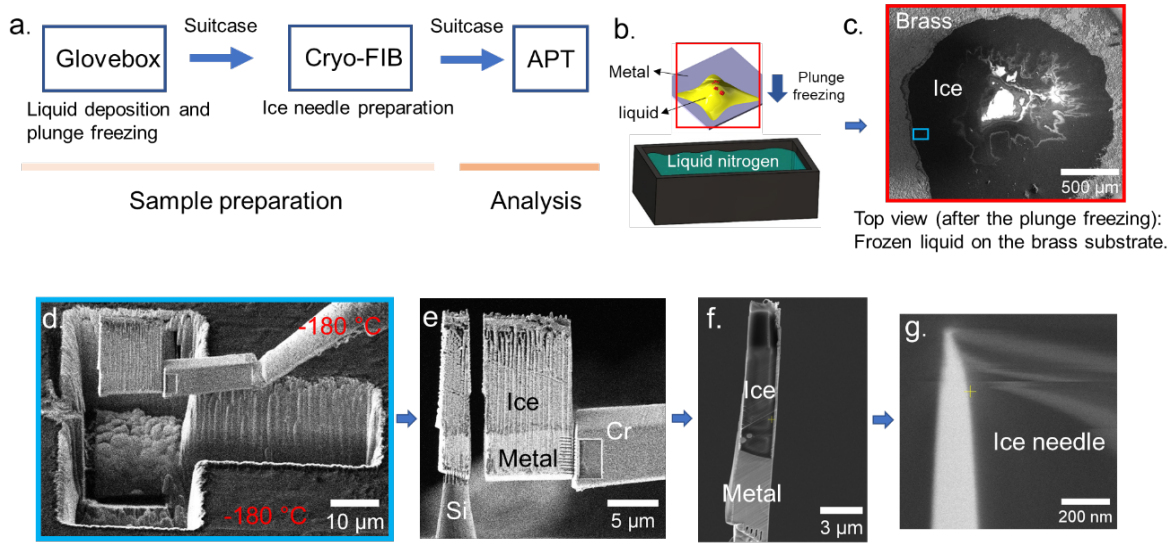

Figure 1: Specimen preparation process: a) Overview of the workflow. b) Schematic representation shows the liquid sample deposited on the metal substrate was plunged frozen into $LN_2$. c) The cryo-SEM image illustrates a thin (6 to 10 µm thick) frozen liquid layer on the metal (brass) substrate, after the plunge freezing. d-e) A lamella, consisting of frozen liquid on top and metal substrate on the bottom, was lifted out and attached to the Si supporting post. f) The ice/metal interface of the lift-out specimen was revealed after the (brass) redeposition layer was removed during FIB milling process. g) The final specimen with an end radius of ~ 80 nm.

### 2.1 General experimental protocol

Figure 1a summarizes the overall instrumental workflow, with methodological details provided in the following sections. We used a solution of ferritin from horse spleen, with a concentration of 10mg/mL in 0.15 M NaCl, sourced from Sigma Aldrich (CAS number 9007-73-2). A drop of the solution was deposited onto a nanoporous brass substrate using a pipette (0.5 µL). The nanoporous brass was prepared by vacuum dealloying at 800˚C for 1h to induce the removal of the Zn into the gas phase as detailed in (Eric V. Woods et al., 2023). This metal substrate was mounted on a sample holder for APT referred to as a puck or cryo-puck. The puck is then immediately plunge frozen into liquid nitrogen, Figure 1b. The plunge-freezing process was operated in a glovebox (Sylatech) filled with nitrogen gas and with moisture control (lower than 10 ppm) to avoid the formation of frost. For more details, please refer to Ref. (Stephenson et al., 2018).

Approximately 1 minute after the bubbling stopped around the cryo-puck, it was transferred into the pre-cooled cryo-transfer suitcase attached to the glove box, subsequently pumped down to ultra-high vacuum (UHV, $<10^{-7}$ mbar). Finally, the suitcase was moved and attached onto a Thermo-Fisher Helios 5 CX Gallium FIB/SEM (Thermo-Fisher Scientific) equipped with an Aquilos-like cryo-stage with free rotation capability and a cryo-cooled Thermo-Fisher EZ-Lift tungsten cryogenic manipulator. The cryo-stage can be cooled down to - 190 °C and the EZ-Lift to - 175 °C by a circulation of $N_2$ gas flow of 190 mg/s through a heat exchanger system within a liquid nitrogen dewar. The difference in the temperature between the stage and micromanipulator is due to the difference in cooling capacity between the two. An electron beam acceleration voltage of 5 kV and an electron beam current of 86 pA were used to image the liquid droplet in Figure 1c and, generally during the specimen preparation process, to avoid damage or melting of the sample.

### 2.2 Cryo-lift-out method

The cryo-LO was performed following the protocol described by (Eric V Woods et al., 2023). At first, a piece of 99.9 % pure Cr (10×5×3 µm) was lifted out and adhered to the tungsten micromanipulator at room temperature, before cooling the sample stage to a set point of - 190 ˚C and the manipulator to - 175 ˚C. Afterwards, the plunge-frozen specimen was transferred to the precooled cryo-FIB chamber using a UHV suitcase (Ferrovac GmbH) also cooled by liquid nitrogen. Inside the SEM/FIB, a lamella of the material of interest (liquid or liquid/substrate) was lifted out and attached to the manipulator via re-deposition welding to the previously attached Cr lamella, Figure 1d. The lamella had dimensions comparable to those used previously, typically 10 x 20 x 3 µm (Eric V Woods et al., 2023) as readily visible in Figure 1d. Therefore, a series of line patterns (30 kV, 40 pA, depth (Si): 0.7 µm) was firstly milled on the interface between the support metal and Cr, to attach the Cr/supporting metal interface via re-deposition (Schreiber et al., 2018; Eric V Woods et al., 2023). To enhance the strength of the junction a rectangular pattern (30 kV, 0.23 nA, 30 s) was used to sputter Cr from the lamella to the interface between Cr and the support metal (Douglas et al., 2023). Finally, the material of interest was lifted out and attached to the Si post on a commercial coupon, using the same procedure of re-deposition welding and cut free. In general, up to 6 such mounts could be made with a single lifted-out lamella, depending on its length.

To shape the specimen, the stage was tilted to face the ion beam, and annular milling was performed with a decreasing ion beam current and voltage. The final polishing step was performed under 5 kV, 15 pA, or 30 kV,7 pA to minimize the Ga implantation, and reduce the extent of the damaged areas. Finally, a needle-shaped specimen with a radius <100 nm could be obtained, as shown in Figure 1f. Strong electron-beam charging is visible near the apex of this ice needle, indicates a poor electrical conductivity of the specimen.

### 2.3   High pressure freezing (HPF)

Pure, deionized (DI) water was vitrified using a high-pressure freezer (*Leica HPM100*, Leica Instruments, Wetzler, Germany). The HPF specimen was then stored in an aluminum container then immersed in a LN2 storage dewar to avoid devitrification, before being transferred onto the cryo-puck. The remaining steps of the workflow are similar to what was described above. The frozen solution was approximately 10µm thick.

### 2.4   Atom probe tomography

After specimen preparation, the cryo-puck was transferred to the atom probe using the same transfer cryo-UHV suitcase system. APT data was acquired on a Cameca LEAP 5000 XS atom probe (Cameca Instruments). The base temperature was - 233 ˚C (40 K), the target detection rate was varied from 0.5% to 0.7%. The data were acquired in the laser-pulsing mode, with a laser pulse energy from 40 to 120 pJ, with a pulse repetition rate of 65 to 200 kHz. Data reconstruction was performed using Cameca's Integrated Visualization and Analysis Software (IVAS) in AP Suite 6.3.1.

## 3   Results

### 3.1   Analysis on vacuum dealloyed nanoporous brass

We used a brass foil dealloyed in vacuum as substrate metal due to its favorable thermal conductivity and hydrophilic surface properties (Eric V Woods et al., 2023). As shown in Figure 1c, the frozen liquid forms a 5–8 µm thick layer on the substrate. A region was targeted for lift-out at the edge of the frozen droplet, where it is thinner and more uniform, and the cooling rate is expected to be faster. The extracted lamella comprised a layer of the frozen liquid and the porous brass substrate. Figure 2a provides a schematic illustration of the expected structure of the final APT specimen divided into three main regions – the liquid layer on top – interface between liquid and porous substrate underneath – and the metallic substrate itself.

Sublimation of water during the transfer process or in the UHV chamber of the atom probe during alignment of the laser or other specimens on the coupon being analyzed can potentially lead to a loss of specimen length prior to the APT analysis, depending on the ice's composition, sublimation point, the actual local pressure and temperature. Yet, based on the phase diagram of pure water (Gerstl et al., 2017), no sublimation is expected at the pressures of $10^{-9} - 10^{-11}$ mbar and temperatures between -230 and -180 °C (40 – 90K) used herein, even if the these cannot be monitored at all times during the transfer process. Differences between specimens are most likely to arise from the sample itself, i.e. the roughness of the initial surface hidden by the frozen droplet that precludes one from precisely visualizing what is being lifted out, but also during the preparation of the specimens as the water mills fast and the sputter yield would depend on the local composition, making it challenging to achieve reproducible specimen shape and length.

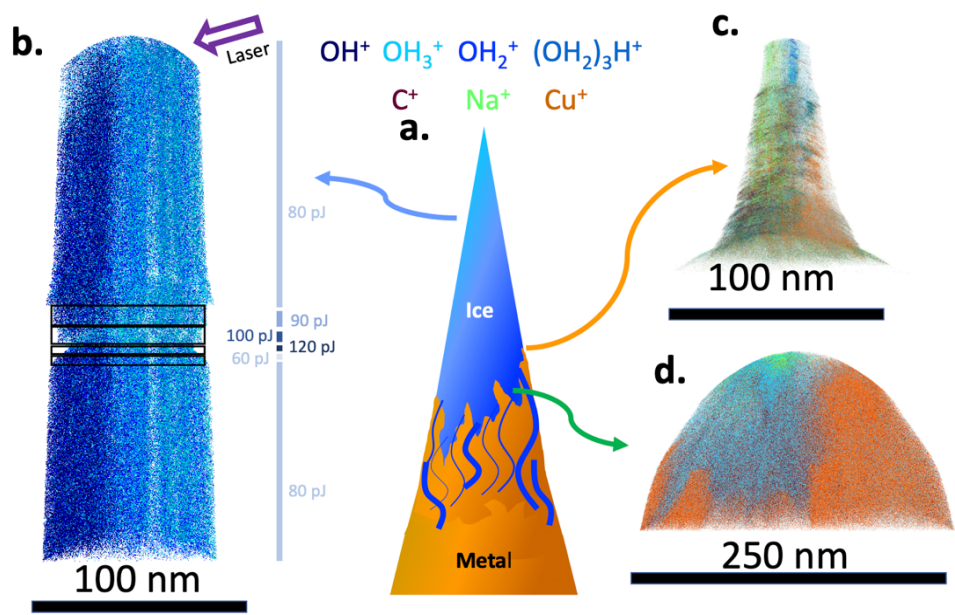

*Figure 2: a) Schematic representation shows the possible micro-fracture locations. b) 3D reconstructions of pure frozen solution with no metal substrate; c) 3D reconstruction of the frozen solution on metal substrate and d) 3D reconstruction with Cu on the edges of the reconstructed dataset*

During the laser alignment and the APT analysis, there are sudden increases in the detection rate (Suppl. Figure 2, Suppl. Figure 3) that can indicate micro-fractures, possibly due to the brittle and maybe porous nature of the specimen, and uneven surface. However, these are not uncommonly strong, with peaks of only up to a factor of 5 times the set detection rate. Reconstructed 3D maps of different APT specimens vary in length, and some include the ice/metal interfaces or only the frozen liquid, as shown in Figure 2b–d. Three distinct configurations were observed:

- Only the frozen solution, not containing the metal substrate, Figure 2b;
- The frozen solution on top of the nanoporous brass substrate, Figure 2c;
- Finally, frozen solution within nanoporous brass substrate on the edges of the reconstructed dataset, Figure 2d.

## 3.2 Frozen solution

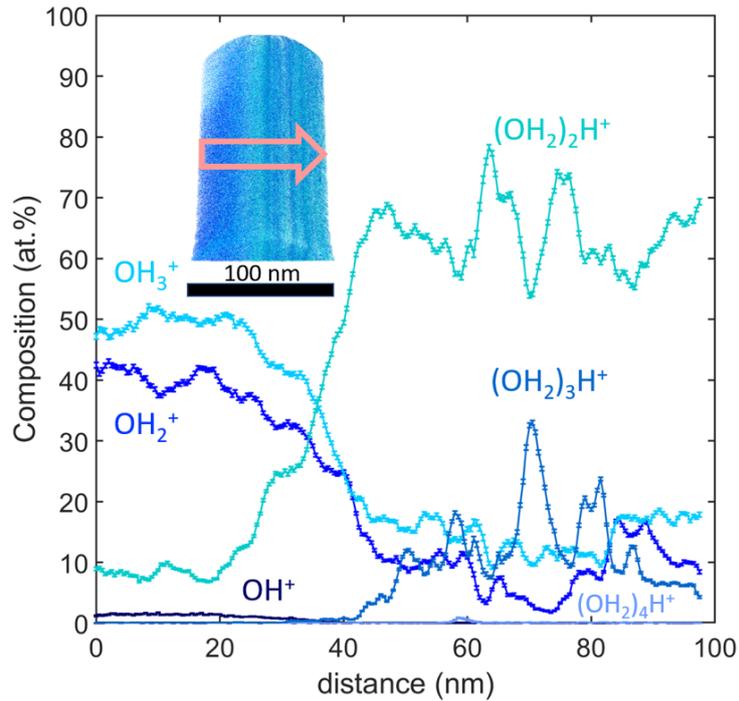

*Figure 3: Ionic composition profile for the most abundant protonated water clusters across the dataset, calculated from the side illuminated by the laser. Inset is shown the selected region-of- interest from the dataset in Figure 2 a. The error bars are calculated based on the counting statistics within each bin of the profile.*

The dataset in Figure 2b contains 69 million ions from a single specimen analyzed at - 233 ˚C (40 K) and across a range of different laser pulse energies (ranging from 60 to 120 pJ) and pulse repetition rates (ranging from 65 to 200 kHz). As expected from previous reports from bulk water specimens (Anway, 2003; Stintz & Panitz, 1992, 1993; Schwarz et al., 2020; El-Zoka et al., 2020), the mass spectrum contains peaks from protonated water clusters $(H_2O)_nH^+$, along with numerous peaks associated with organic compounds, and dissolved metal ions.

Figure 3 shows the varying ion distributions of $OH^+$ and $(H_2O)_{1-4}H^+$ obtained for a section of the long dataset with 80pJ and 200 kHz in Figure 2b. The smaller cluster chains, such as $OH^+$, are detected preferentially on the side away from the laser incidence, sometimes referred to as the shadow side (Bachhav et al., 2011; Schwarz et al., 2022). Longer protonated water chains are detected on the laser side. The higher temperature reached on the laser incidence side and the poor thermal conductivity of water lead to preferential field evaporation that increases the local specimen radius and hence decreases the local electrostatic field. Consequently, the shadow side is under a higher electrostatic field condition to reach the same detection rate as the laser incidence side, as previously reported for a range of other materials (Sha et al., 2008; Müller et al., 2012). The higher temperature can help drive the formation of the larger protonated water clusters (Schwarz et al., 2020), while a higher electrostatic field leads to the more likely dissociation or fragmentation of these larger protonated water clusters such as $(H_2O)_{2-3}H^+$ into smaller molecules such as $H_{1-3}O^+$ on the shadow side. The concentration of $OH^+$ remains generally low, but the small increase observed on the shadow side is likely because of the decomposition of $H_2O^+$ with increasing electrostatic field. These results broadly agree with previous reports (Schwarz et al., 2020). Density functional theory (DFT) calculations on bulk water samples suggest that the formation of $OH^+$ is unlikely, as the pronated water clusters are thermodynamically more stable (Segreto et al., 2022).

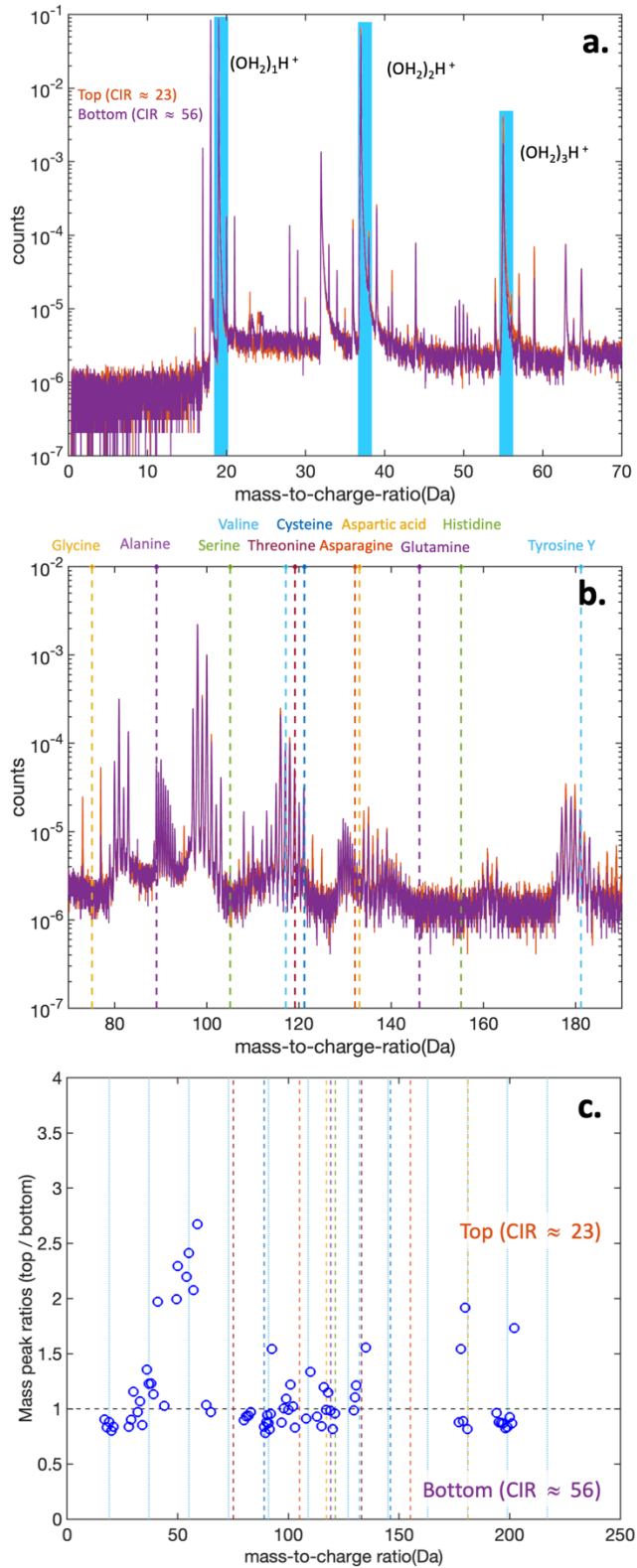

*Figure 4: a) Section of the mass spectrum from 0 to 70 Da, normalized to the total counts in the data set, for two different regions obtained at the same laser pulse energy (80 pJ) in Figure 2, with the top part of the dataset plotted in orange and the bottom part in purple. Peaks for the three smaller protonated water clusters are marked in blue. b) section of the same mass spectra in the range 70 – 190 Da. The vertical dashed lines mark the mass of complete amino acids typically found in the subunits of the ferritin. The light blue dotted lines indicate the peak position for water clusters $(H_2O)_nH^+$. The color-coding of the other vertical dashed lines is the same as in (b). (c) ratio of the amplitude of 70 peaks found in common between the two mass spectra shown in (a) and (b), with indicated the cluster-ion ratio (CIR) for the two sections.*

Two $10^7$-ions sections were extracted respectively at the top and bottom of the dataset displayed in Figure 2 b. Both sections were analyzed at 80 pJ. The corresponding mass spectra are plotted in Figure 4a-b in orange (top) and purple (bottom) respectively. The mass spectra are normalized to the total count within the dataset. In the part of the mass spectrum in the range 0 – 70 Da, the peaks pertaining to the three most abundant protonated water clusters (H$_2$O)$_n$H with n = 1–3 are displayed and marked with blue rectangles in Figure 4a. One can expect that most of the other peaks in the mass spectra in Figure 4a and Figure 4b are related to fragments from the polypeptide chains initially surrounding the ferritin core, as normally the solution only contains NaCl. This indicates that the overall structure of the ferritin has been lost, most likely during the freezing process due to the ice crystal formation.

The relative intensities of the peaks for the water clusters has previously been suggested to reflect the electrostatic field conditions during the analysis (Woods et al., 2025; Schwarz et al., 2020). Woods et al. define the cluster ion ratio (CIR) as $\sum (H_2O)_{n=1,2,3}H^+ / (H_2O)_3 H^+$, as an equivalent to a charge-state-ratio to facilitate comparison across datasets (Woods et al., 2025). The CIR for different the sections of the dataset in Figure 2b are plotted in Suppl. Figure 1 along with all the data from (Woods et al., 2025), and shows a consistent trend also for this ferritin solution.

In addition, the evolution of the applied voltage, laser pulse energy, laser pulsing frequency and background for the dataset in Figure 2 is plotted in Suppl. Figure 4. At a constant laser pulse energy, the voltage increase is relatively steady, as expected during the APT analysis, to compensate for the progressive blunting of the specimen. The level of background computed by the commercial software comes from the baseline level of the time-of-flight (ToF) spectrum and is hence expressed in relative counts per ns. This is then converted into a number of counts per unit of mass-to-charge ratio using the usual conversion (Larson et al., 2013). In the calculation of the mass spectrum, it is possible to use a quadratically increasing bin size to calculate the histogram, allowing for estimating the baseline level as shown in Suppl. Figure 5. Overall, the background decreases very slightly with increasing laser pulse energy from 80 to 120 pJ, and reaches a maximum at 60pJ. The increase in laser pulse frequency from 60 – 200 kHz leads to only a moderate increase of the level of background from approx. 8 to approx. 11 ppm.ns$^{-1}$. These trends are consistent with previous reports (Schwarz et al., 2020; El-Zoka et al., 2020). In the end, the differences here are rather limited as readily visible in Suppl. Figure 5.

When considering the identification of the peaks , colored vertical lines were added in Figure 4b at masses of some of the identified amino acids found in the protein shell around the ferritin core (Hempstead et al., 1997). Details pertaining to these amino acids are reported in Suppl. Table 1. There is a possibility that series of peaks in the range 115–121 Da could be related to cysteine or threonine or valine that have lost one or more hydrogen atoms. There are similar sets of peaks up to 250 Da and hence above the mass of the heaviest individual amino acid identified, suggesting that combinations of individual amino acids are coming off still assembled together. This is not to suggest that complete individual amino acids could always be readily expected, more to insist on the fact that the detected fragments can be found from combinations of multiple amino acids.

Finally, the two sections of the dataset in Figure 4a and Figure 4b were not acquired under the exact same electrostatic field conditions, CIR of approx. 23 and 56 in the top and bottom section respectively. Yet peaks appear rather comparable in position as indicated in Suppl. Figure 6, in which peaks were identified based on their prominence with respect of the values of their nearest neighbors. The ratio of the amplitude of the 70 peaks found in common between the two mass spectra is plotted in Figure 4c. This ratio is very close to unity in most cases. The higher ratio for the peak at 55 Da indicates a higher relative abundance of this peak in the top section of the dataset in which the CIR and hence the electrostatic field is lower. The peaks pertaining to the heavier cluster ions appear more heavily concentrated in the bottom section (ratio <1). This could appear counterintuitive, as one could expect the larger cluster ions to dissociate under the higher electrostatic field conditions (Dietrich et al., 2025). However, it has previously been reported that solutions containing organics frozen in similar conditions were far from homogenous (Woods et al., 2025). Here, the ionic ratios for water clusters vs. organic species identified in the mass spectra, shows that the top section contains approx. 4.7% organics vs. 2.9% in the bottom section, indicating a heterogenous distribution within the solution itself. Further studies will be needed on simpler systems, to ensure that these can be undoubtedly identified. Nishikawa and co-workers reported on individual amino acids deposited on metallic needles analyzed by APT (Nishikawa et al., 2011; Nishikawa & Taniguchi, 2017). These results cannot be readily used herein as the electrostatic field was much higher in their experiments that did not have the aqueous matrix.

## 3.3 Ice/metal interface

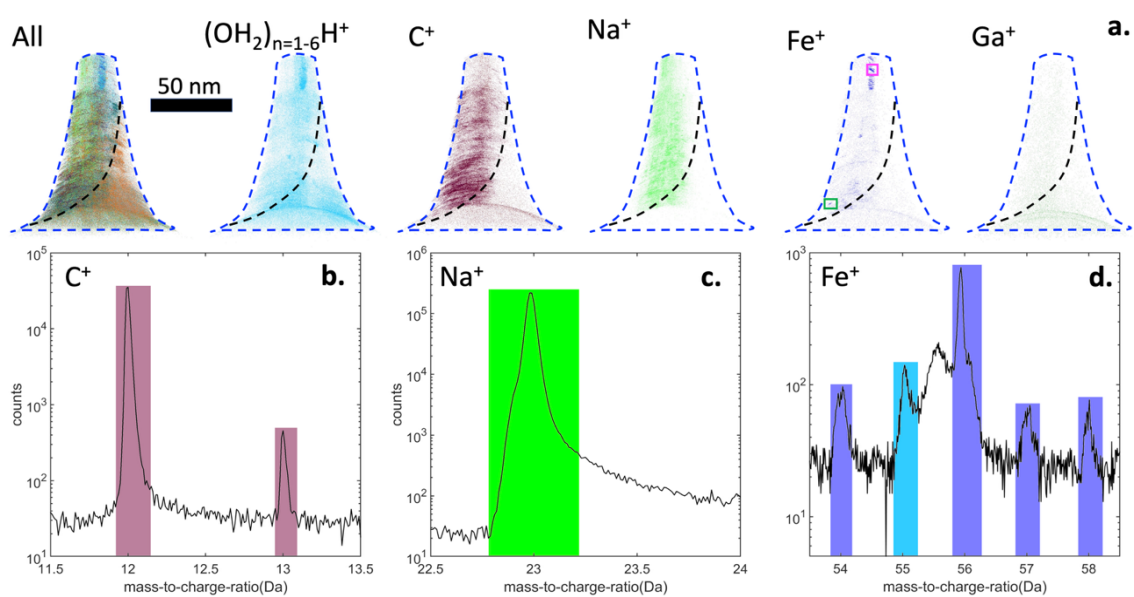

*Figure 5: a) Distribution of selected identified elemental and molecular ions. b–d) Section of the mass spectrum for the peaks pertaining to $C^+$, $Na^+$, and $Fe^+$.*

The data in Figure 2c from the ice-metal interface is detailed in Figure 5. Datasets acquired from specimens containing this interface are in general shorter compared to measurement of the solution. Here again, the specimen fractured early during the APT acquisition, near the ice/metal interface, marked by the black dashed line. One reason for the shorter measurement may be the large difference in field strength between frozen water – 8-10 V/nm (Segreto et al., 2022) – and brass – 30-33 V/nm (Tsong, 1978). Considering the complexity of the mass spectrum, which is very similar to Figure 4, not all peaks could be identified and we focus the data analysis here on the distribution of a selected few "key elements": $Na^+$, $Fe^+$, and $C^+$, and the combination of the protonated water clusters, displayed in Figure 5a. Figure 5b–d plots the peaks in the mass spectrum corresponding to $C^+$, $Na^+$, and $Fe^+$. It should be noted that Cu and Zn are detected as $Cu^+$, $Zn^+$, as well as hydrated molecular ions, i.e. $CuH_2O^+$, as previously noted by Woods et al. (Woods et al., 2025). Section 4.1 below provides further discussion on metal-containing molecular ions.

Compared to the dataset containing only the solution, Figure 3, the content of the larger protonated water clusters containing $(H_2O)_{n=3–5}H^+$ is relatively low, which can be attributed to their dissociation into smaller cluster ions such as $OH^+$ and $H_2O^+$ that are observed near the interface with the metal.

This can be expected, as the metallic surface is at the full electrostatic potential applied to the specimen holder, whereas it will be lower at the apex of an ice needle, as ice is in generally a poor electron conductor. A resulting higher electrostatic field near the interface favors fragmentation and dissociations into smaller clusters as previously reported (Schwarz et al., 2020; El-Zoka et al., 2020).

In addition, a higher concentration of solutes was detected near the ice/metal interface, notably, the distribution of $Na^+$ mirrors that of $H_{1-3}O^+$ ions, suggesting a common origin from the liquid. Overall, the $Na^+$ concentration near the interface is significantly higher than that observed for $H_{1-3}O^+$, which suggests that the solution at the ice/metal interface was highly concentrated in Na, as already discussed in (El-Zoka et al., 2020). This can be attributed to the possible interaction of salt ions with the brass substrate prior to freezing, but also the slow freezing process can lead to the rejection of solutes into the liquid as their solubility in the solid is extremely low (de Almeida Ribeiro et al., 2021). This means that ions can be transported during solidification very far from where they were initially within the solution. From the ion distribution map, the $C^+$ ions are distributed throughout the water ice matrix. Here again, this carbon likely originates from the protein shell of the ferritin particles, and the presence of the $C^+$ ions can be explained by the dissociations of larger amino acids molecules caused by the intense electrostatic field near the interface between the frozen solution and the metal substrate.

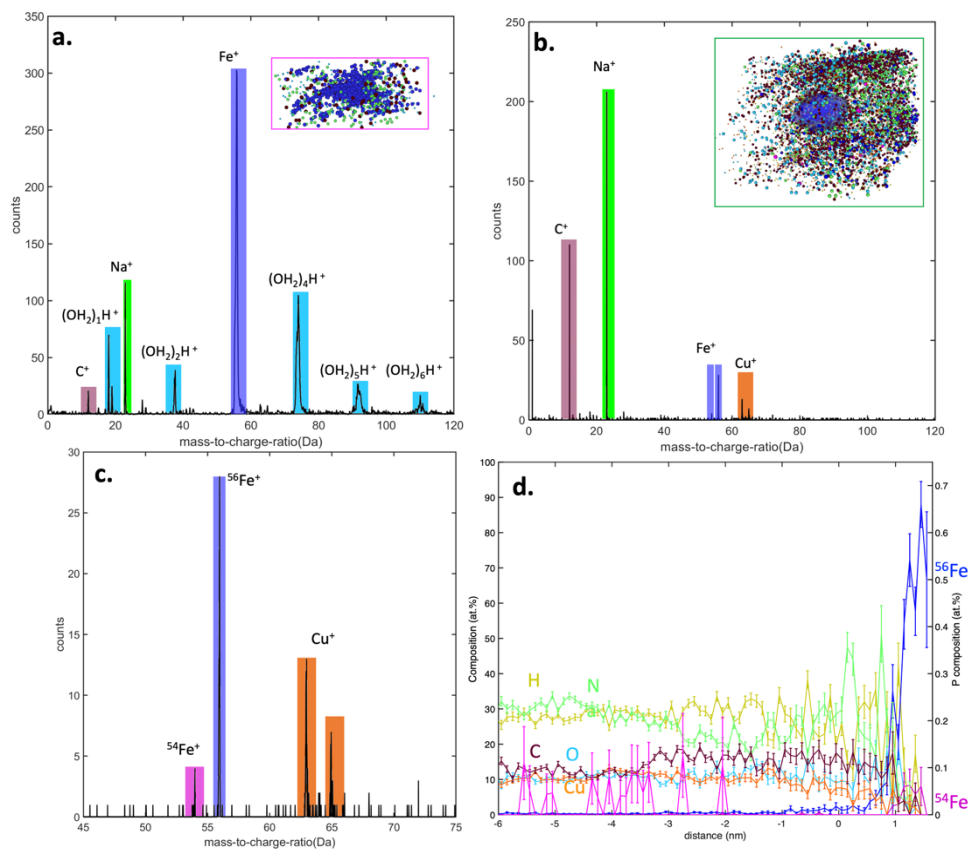

*Figure 6: a) mass spectrum in a cuboidal region-of-interest centered on a cluster of ions identified with a mass-to-charge ratio of 56 Da. Inset is the corresponding cluster extracted from the 3D reconstruction; b) similar analysis with the mass spectrum of a cluster shown inset, selected where the spatial distributions of the peaks at 54 and 56 Da overlap; c) section of the mass spectrum extracted from the inner side of this same isosurface in the inset in b), and (d) proximity histogram calculated from the corresponding isoconcentration surface.*

Ions from the peaks labelled as Fe appear agglomerated in several regions across the dataset displayed in Figure 5a. One, delineated by a pink box, was extracted and the corresponding mass spectrum is plotted in Figure 6a (note the linear scale of the mass spectrum). The main peaks are identified. Protonated water clusters $(H_2O)_nH^+$, with n up to 6, are all detected with the notable exception of the $(H_2O)_3H^+$ at 55 Da. A strong peak is detected for $Na^+$, already noted by (Perea et al., 2016), however conversely to their results, no P-containing peaks and only a negligible amount of C-containing species are detected. One should note though that the peak marked as $Fe^+$ at 56 Da is not accompanied here, by the peaks pertaining to the other isotopes of Fe, particularly at 54 Da, conversely to what can be seen in Figure 5d. This hence indicates that the peak was unfortunately misidentified. $C_2O_2^+$ or $C_2NOH^+$ are molecular ions with the same mass-to-charge at 56 Da.

Examining the spatial distributions of the peaks at 56 and 54 Da (Suppl. Figure 7) reveals some overlaps, but also differences. A region-of-interest, marked with a green box in Figure 5a, was extracted where the distributions overlap. The mass spectrum is shown in Figure 6b, and the particle is highlighted by the blue isoconcentration surface in the inset of the figure. A section of the mass spectrum from inside this isoconcentration plotted in Figure 6c evidences the presence of peaks pertaining to the two most abundant isotopes of Fe. Their ratio is close to the natural abundances, 8.9% and 90% respectively for $^{54}Fe$ and $^{56}Fe$. A composition profile was calculated in the form of a proximity histogram (Hellman et al., 2000) is plotted in Figure 6c, including for these two isotopes alongside the main other constituents, namely C, O, H, Cu and Na. The particle contains almost 80 at% Fe, surrounded by a shell enriched in C. Conversely to results from (Perea et al., 2016), no notable increase in P can be observed, with no peak above background for $FePO_2^+$ (119Da), and the peak for $PO_2^+$ (63 Da) overlapping with the most abundant Cu isotope. It is noteworthy that the composition in Na is nearly 30 at% around the particle. Below will be discussed how these observations support the hypothesis that the original ferritin particles were damaged or even destroyed by the crystallization and formation of ice crystals during freezing, and the destroyed protein shell ended up distributed across the solution.

### 3.4 High-pressure freezing

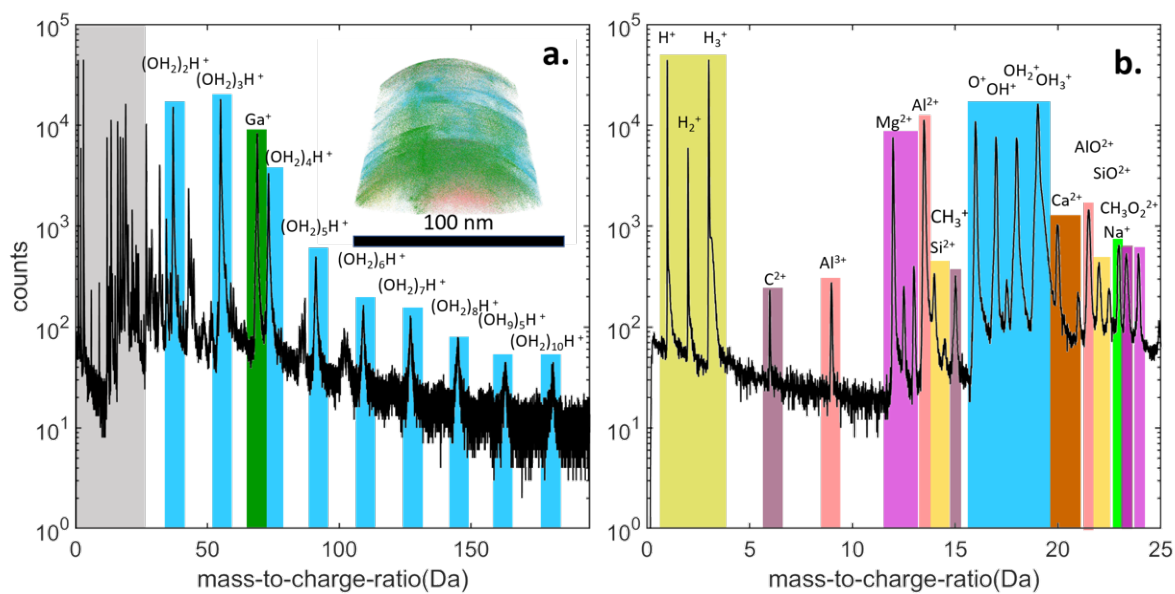

Figure 7: The mass spectrum from the HPF-prepared water on an Al planchette plotted logarithmically for two different ranges of mass-to-charge-ratios, a) 0–200 Da and b) 0–25 Da. Inset in a) is the corresponding 3D tomographic reconstruction. In both the mass spectra, the corresponding ranges are marked in colors and peaks are labelled.

Preliminary results demonstrate the feasibility of analyzing frozen specimens fabricated from HPF specimens. An aluminum canister containing the HPF ice planchette was transported into the glovebox in a LN2 dewar, and the canister was opened within a liquid nitrogen bath. The HPF planchette was then mounted onto a clip-holder on a cryo-puck, and transferred into the pre-cooled vacuum suitcase. Needle-shaped specimens were then prepared using the same cryo-LO protocol, Suppl. Figure 8, and transferred to the APT analysis chamber using the same cryo-suitcase. Compared to the crystalline ice samples shown in Figure 1 and Figure 2, the HPFed sample is more challenging to prepare and transfer, as the vitreous-to-crystal phase transition is irreversible, and will happen at a temperature above -135 °C (Marko et al., 2006). In principle, the temperature of the specimen has remained below 90 K (-183 °C) during the whole transfer process. As a result, the amorphous state of the ice should be maintained (Stintz & Panitz, 1993), even if, here, the vitreous state of our samples could not be confirmed.

Figure 7a–b shows the mass spectrum from one of the acquired datasets, with Figure 7b corresponding to the region marked in grey in Figure 7a. The main peaks in the mass spectrum were identified and marked. Compared to the data from the frozen solution shown in Figure 4, which reveals protonated water clusters ranging from $(H_2O)_{n=1-3}H^+$, the data from the HPF specimen contains protonated water clusters up to $n$ = 12. A similar behavior was previously reported for thin layers of amorphous ice deposited on a pre-sharpened tip (Stintz & Panitz, 1993; Dirks et al., 1992). In addition, the data contains peaks of $Al^{1-3+}$, $Mg^{1-2+}$, and $Si^{2+}$. These detected elements belong to the Al planchette, made of a commercial EN AW-AlMgSi alloy, referred to as 3.2315 or EN AW 6082. In Figure 8 the maps of the individual elements and molecular ions are plotted, showing that these elements are primarily located at the bottom of the dataset, confirming that the analysis has reached the interface between the frozen water and the metal, marked with a dashed black line, just before the fracture of the specimen. The surface contamination with C is also visible, which could come from exposure to the environment or cleaning with alcohol-containing solutions. Ca and Na may be residual from the DI water or left from the drying of water or a solution used to clean the planchette.

The distribution of Ga is very heterogenous. A blue and a pink arrow in Figure 8 indicate where the composition profiles were calculated. Figure 9a plots the composition profile along the blue arrow, across the solution-metal interface, and shows segregation of Ga at the liquid-metal interface. This can be related to the change in sputter yield of Ga for different materials, which roughly scales inversely with melting point (Prenitzer et al., 2003). Al is much harder compared to ice, therefore the implantation of Ga ions is reduced. Figure 9b shows the composition profile along the pink arrow that crosses three individual Ga-rich clusters. Within the clusters, the Ga reaches nearly 100 at%. This suggests a high degree of implantation and associated damage during preparation. For the sample preparation, rectangular cuts were performed on the four sides of lamella during the shaping process to remove the redeposited layers following annular milling process. This increased the time during which annular milling was used to prepare the needle-shaped specimen, however, using the rectangular pattern in respect to the annular milling can cause deeper penetration of the Ga ions, even at cryogenic temperature (Chang et al., 2019; Prenitzer et al., 2003). A second dataset prepared from the same lamella, also fractured at the liquid-metal interface, also shows a slight increase in Ga at the interface.

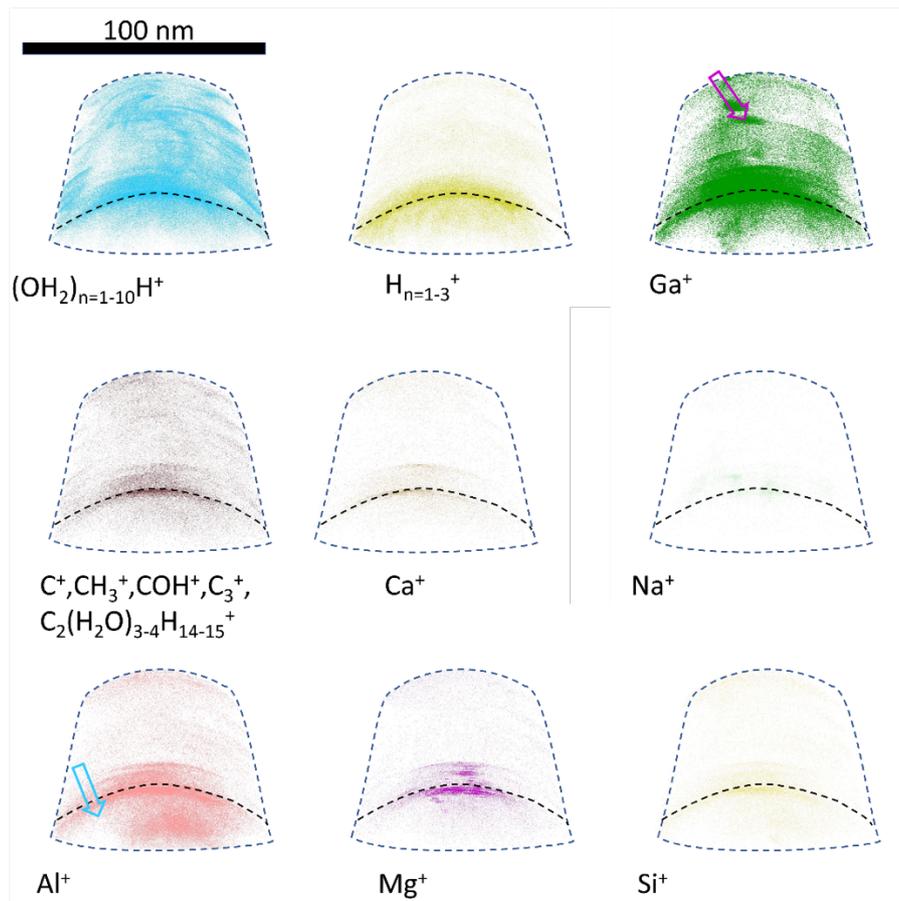

*Figure 8: 3D reconstruction elemental map for different elements and molecular ions in the HPFed water specimen. The dashed blue line outlines the edges of the dataset, while the dashed black line roughly indicates the position of the metal-liquid interface. The blue and pink arrow indicate the position of the composition profils displayed in Figure 9.*

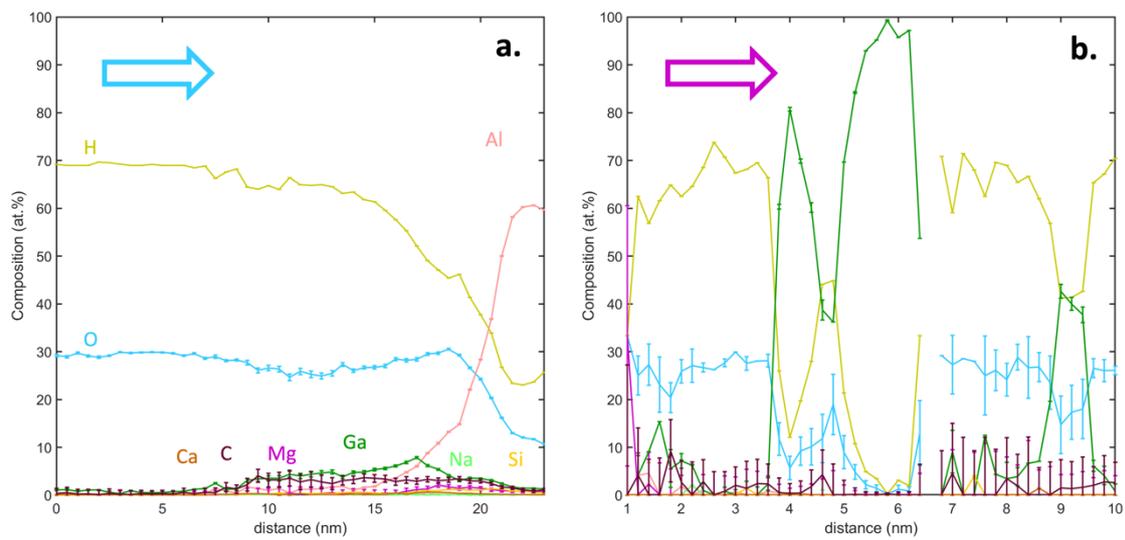

*Figure 9: Representative composition profiles across (a) the solution-metal interface and (b) Ga-clusters.*

# 4 Discussion

## 4.1 Consideration on the substrate

The selection of the substrate metal that supports adhesion of the liquid sample is crucial for specimen preparation and data acquisition. An ideal candidate for the substrate metal should have the following qualities:

1. Good thermal diffusivity to ensure rapid cooling rate to minimize the damage from the crystallization to the biological structure of ferritin. In addition, the thermal diffusivity of the substrate material also plays a significant role in reducing thermal effects when using laser pulsing in APT.
2. Sufficient strength: The metal strength should be strong enough to survive the high field in the atom probe. Secondly, the metal should support adhesion using re-deposition method with a sufficient strength at the junction.
3. Surface of the substrate: The surface of the substrate is essential for the adhesion/fixation of the liquid phase. Therefore, different approaches have been used to provide rough or nanoporous materials.
4. Controlled hydrophilicity: the flat metal sheet should be able to accommodate the liquid sample and form a thin liquid layer (5-10 μm thickness), allowing the lift-out of a lamella containing frozen-liquid (5-7 μm thickness) on the top and metal support on the bottom. A hydrophilic surface allows the liquid to spread on the surface evenly, and not form a bubble-shaped (thick) liquid drop.
5. Simple composition: each component of the alloy could add peaks to the APT mass spectrum and will make it more challenging to interpret unambiguously the data due to potential overlaps. In addition, the metal should be inert as it can corrode and dissolve in the liquid and be encapsulated in the ice needle. The selected alloy should, in an ideal case, be corrosion resistant in the environment of the solution and if metal ions get dissolved in the solution it should not contain elements normally found in the solution (for example, iron in this study).
6. Milling rate: Considering the specimen preparation time during LO, the milling rate of the metal should be taken into consideration, to reduce the sample preparation time.

Here, we reported results on dealloyed brass, but other metal substrates have been tried, including nanoporous gold (NPG), magnesium and aluminum. These had limited successes, for different reasons laid out in the following.

Interfaces between low- and high-evaporation-field phases in APT are known to be challenging because of possible delamination associated to stress concentration at the interface (Rolland et al., 2015; Kölling & Vandervorst, 2009), and are locations of intense trajectory aberrations due to local magnification effects that distort the 3D reconstruction (Vurpillot et al., 2004; Larson et al., 2011, 2012). Additionally, organic compounds, as well as iron and sodium ions, tend to concentrate near the liquid/metal interface, changing the liquid composition, and hence likely its field evaporation behavior, but also making this a region particularly important to characterize to understand physical processes at the liquid/metal interface.

This forces us to consider the critical role of the interface between the solution of interest and the substrate, and particularly its microstructure (Stender et al., 2022). For this reason, nanoporous gold was shown to be a promising candidate (El-Zoka et al., 2020), as its large surface area provides an extensive ice/metal interface, resulting in a high strength of the interface, but also avoids an abrupt transition between the solution and the substrate. Moreover, Au, and Ag to a lesser extent, are less reactive and less likely to dissolve in water, which could simplify data analysis by reducing complexity. However, here, we could not obtain satisfactory data, Suppl. Figure 9.

A recent study has demonstrated the formation of a porous hydroxide at the interface during aqueous corrosion of Mg (Schwarz et al., 2024). Compared to brass, Mg is mechanically softer and has hence a higher FIB-milling rate, closer to that of ice, making is potentially simpler to obtain specimens with a smooth surface. These required adjustments of the parameters used for FIB specimen preparation, but we tried to use a similar Mg flat substrate for liquid deposition, leaving the ferritin solution on the Mg surface for 30 mins for corrosion and to form of the porous network. The sample was then plunge frozen, and transferred into the FIB for cryo-LO and sharpening. However, samples made on corroded-Mg fractured early, resulting in the collection of limited datasets not worth reporting herein. This may be caused by a different corrosion mechanism, since it is known that the electrolyte composition and the presence of organic material, change the corrosion mechanism for Mg. The saline buffer solution containing ferritin may not have caused the formation of a similarly dense porous network as the pure water had in the reported study (Schwarz et al. 2024). We therefore faced common delamination of the hydroxide-metal interface, with the porous network not providing sufficient strength to the interface compared to the nanoporous metals.

For the specimen prepared from HPF deionized water, the substrate metal of the planchette was a commercial Al alloy. An advantage is that the concentration of Al away from the interface appears to be low, suggesting that the substrate was rather inert, which in turn limits the adhesion to this flat substrate, which was not as high as with the brass substrate. The difference in milling rate between the Al alloy and the ice layer resulting into the formation of a shoulder at the interface, Suppl. Figure 8, that likely facilitated the fracture at the interface, due to the increase of the base voltage to maintain the target detection rate. This has limited the amount of data collected. In the future, the approach for specimen preparation will need to be adjusted to improve the success rate for these samples, or even to prepare LO without the underlying Al planchette substrate (Eric V Woods et al., 2023).

Dealloyed brass allowed for acquiring data routinely. However, some metal ions dissolve in the liquid and concentrate near the metal/liquid interface, and their concentration decreases with increasing distance from the bulk metal – i.e. they were not detected in the dataset displayed in Figure 2b. There is in principle no liquid residue from the dealloying in this case, as the sample was dealloyed in vacuum and not through a chemical route, as for the previously reported nanoporous gold (El-Zoka et al., 2020). Yet the sample surface was subjected to several rounds of cleaning with deionized water, which may have caused some dissolution or corrosion already as suggested by recent results obtained on pure Cu (Kang et al., 2024). In the future, the dealloyed brass could be used as a template for depositing an inert material for instance.

In addition, it has been shown that the amino acids form complexes with dissolved Cu ions both in solution and at the interface with the Cu. This occurred despite the fact that during the sample preparation process, the ferritin solution was on the metal surface only for 3–5 seconds at room temperature before being plunged into the liquid nitrogen. However, the metal appears to react with the saline/buffer solution at the liquid/metal interface. Closer to the interface, the concentration of metal-containing molecular ions increases, including Cu- and Zn- containing ions with S and Cl as well as with C, N and O as evidenced in Figure 10a and b respectively. As discussed in (Woods et al., 2025), Cu is known to bind to amino acids in solution, so this finding is not surprising. However, it can affect our capacity to identify peaks from the sampled material. In the future, this could help to fix the proteins inside the nanopores of brass (Woods et al., 2025) or within NPG (E. V. Woods et al., 2023), but this will need to be studied in more details.

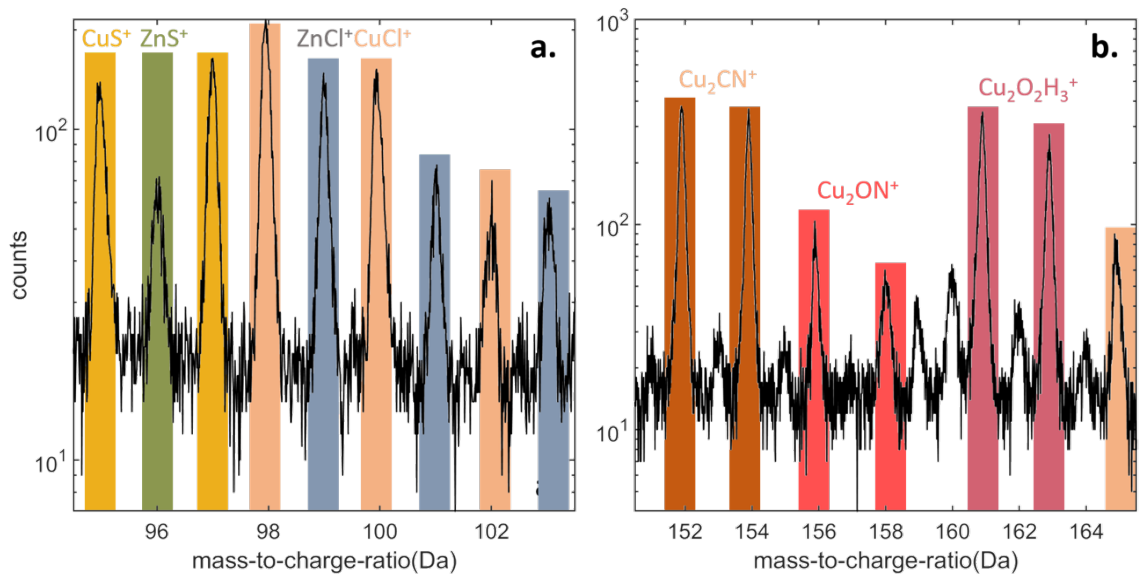

*Figure 10: Mass spectrum from the solution-metal substrate for brass showing a) $CuS^+$ and $ZnS^+$ ions, as well as $CuCl^+$ and $ZnCl^+$ and b) $Cu_2CN^+$, $Cu_2ON^+$, and $Cu_2O_2H_3^+$.*

## 4.2 Substrate geometry

The data discussed so far has been acquired from LO from flat substrates. Specimens were also prepared on a copper TEM half-grid (PELCO FIB Lift-Out TEM Grids, *Ted Pella Inc., Redding, USA*), as shown in Suppl. Figure 10. Specimen preparation on the half-grid is more time-consuming because the surrounding material needs to be milled away to provide the necessary clearance to enable APT analysis, of approx. 100 µm. This necessitates the removal of a large amount of ice, not unlike what had been proposed by Schwarz et al. (Schwarz et al., 2020), or what was discussed extensively by (Eric V. Woods et al., 2023). This approach was efficiently used in the past on a cryo-plasma FIB (El-Zoka et al., 2020), but it is very time consuming on a Ga-FIB, i.e. 6–8h per specimen. In contrast, with the cryo-LO method, we can prepare 5 to 7 needles in a similar amount of time.

Ultimately, the cryo-LO can be performed also directly from the ice (Eric V Woods et al., 2023), and the bonding through re-deposition complemented by *in-situ* sputtering from the micromanipulator or a metallic target provides sufficient strength to the interface to perform APT. This may well be a preferred route to avoid possible failures and complexities in the field evaporation at the substrate-solution interface.

## 4.3 Specimen preparation and transfer

There are also questions as to whether the ion impact during specimen preparation in the cryo-FIB does not lead to substantial damage and heating that could lead to crystallization of the amorphous ice, i.e a temperature rise above -135 °C from the -190 °C of the stage. There are ample evidence from the literature of using cryo-TEM on lamellae produced by cryo-FIB and that demonstrated that the vitreous state is maintained (Parmenter & Nizamudeen, 2021; Schiøtz et al., 2024; Schaffer et al., 2019). Comparable currents were used for their preparation as for ours, and despite the difference in specimen geometry, we expect that there was no heating sufficient to cause significant damage during the preparation of the APT specimens.

Implantation of ions from the FIB was evidenced in Figure 8 and Figure 9, with a strong accumulation of Ga at the interface with the metal and in the form of Ga-rich platelets. These energetic ions could have caused substantial damage to any particles in the solution. In the data for the crystalline ice, displayed in Figure 2 to Figure 6 and discussed in the corresponding sections (3.1, 3.2, 3.3), no such Ga segregation was observed. This could be ascribed potentially to a different behavior of the crystalline and vitreous ice, or maybe to a higher degree of control during the specimen preparation.

Finally, as stated above, we assume that the specimen is maintained through the entire preparation and transfer below the temperature that would cause significant sublimation and / or crystallization. As a result, the vitreous state of our samples should be maintained. Using electron diffraction in the TEM could provide a final proof that the sample has maintained its vitreous state. For now, however, our experimental infrastructure does now allow for transferring the specimen into a TEM at cryogenic temperature and under a protected atmosphere to avoid frost formation on the APT specimen. In addition, electron illumination of the ice can lead to extreme damage to the specimen that could preclude subsequent APT analysis as reported for the analysis of vitreous ice (Dubochet et al., 1982; Talmon et al., 1986).

### 4.4 Compositional analysis

Generally, the mass spectrum in all measurements reveals the presence of Na, Fe, C, protonated water clusters $(H_2O)_nH^+$, numerous organic fragments, and dissolved metal ions from the substrate itself, as discussed in more details below. Analyzing the evaporation and fragmentation behavior of the matrix, in the case of pure water, is already challenging (Stender et al., 2022; Stintz & Panitz, 1992, 1993), and the addition of solvated species further complicates the analysis (Schwarz et al., 2021, 2022). Unambiguously identifying all peaks proved challenging, as many organic fragments e.g. $C_2H_4^+$ (28.06 Da), $CNH_2^+$ (28.03 Da) and $CO^+$ (28.01 Da) have the mass-to-charge state ratios near 28 Da and the needed mass resolution to separate these peaks (in the range of ΔM/M of 1100 ) is higher than the resolution reached in our APT analyses (typically 750 full-width-half-maximum and 350 full-width-tenth-maximum). This is just an example, but these possible overlaps extend across the full range of measured mass-to-charge ratios.

It was sometimes discussed that APT can be used "standardless", and did not need calibrations as is the case for other microanalysis techniques. Yet analyzed samples are compositionally and structurally ever more complex, and often there is not yet a complete theoretical understanding of the field evaporation process for e.g. organics. Earlier theoretical work (Wang et al., 2006) and sparse experimental work (Nishikawa & Kato, 1986; Nishikawa et al., 2011, 2009; Gault et al., 2010; Eder et al., 2017; Stoffers et al., 2012; Meng et al., 2022) have demonstrated that even simple carbon chains can dissociate into numerous fragments under the extreme electrostatic field conditions of an APT experiment. Several authors recently reported using a "standard" e.g. when analyzing a glucose solution, Schwarz et al. also reported analysis of solid glucose to try and better understand the origins of some of the observed peaks (Schwarz et al., 2021), and followed a similar approach for the analysis of bone recently (Schwarz et al., 2025) to support the identification of peaks as originating from collagen. Woods et al. tried to perform analysis of individual amino acids in solution, but even then the changes in the amplitude of the mass peaks of the different fragments as a function of the electrostatic field conditions made it challenging to achieve reproducibility (Woods et al., 2025).

A possibility could be to analyze targeted individual amino acids, and build a database of possible fragments as a function of the field conditions. In parallel, this will help advance the understanding of the underlying principles of the fragmentation process and devise models to either predict possible fragments or assess the identity of a fragment based on a known biomacromolecule sequence. There are interesting efforts in that direction (Segreto et al., 2022; Dietrich et al., 2020, 2025) on relatively simple molecules already, more will be necessary in the future. There will also be a need to develop data process techniques that offer ways to reconstruct these chains from the incomplete set of information that the APT provides, due to the detector efficiency. Sequential processing of ion impacts arriving spatially in proximity to each other may help, but this will have to be developed and tested on simpler systems. Finally, once these fragments are identified and potentially locally reconstructed, this information will need to be reconciled with the general 3D reconstruction. At this stage, and because the atomic volume is used for the depth incrementation (Gault et al., 2011; Larson et al., 2013), issues with mass peak identification necessarily leads to artefacts.

### 4.5   Damage from freezing and the need for vitrification

In cases where the freezing process is too slow, the ferritin particles are likely to remain within the liquid and be carried towards the water-metal interface. The remaining liquid will see a drastic composition change. For instance, in the data reported in Figure 6 c, the Na composition can be up to approx. 30 at% near the interface, indicative of a brine with a completely different solidification behavior (Wolfe & Bryant, 1999). There are many reports in the literature of unfolding and destruction of the conformation of proteins (Crowe et al., 1998), along with their oxidations, driven by the change in the composition of the solution, and as crystallization occurs (see point above) and the formation of ice crystals (Tan et al., 2021). These cause a substantial volume expansion (9% for hexagonal ice) leading to stresses on the proteins and subsequent damage (Rusciano et al., 2017). The change in the protein folding and conformation in the case of the ferritin could leave the Fe-core exposed and potentially to leave the protein shell, and the core itself can potentially also dissolve. The fact that Fe appears to be primarily detected close to the water-metal interface, i.e. where the last ice will form and with a solution's composition completely modified, makes damage to the ferritin structure extremely likely, possibly to the point of complete destruction of the particles themselves.

Using vitreous ice for specimen preparation is essential to preserve the near-native state of biological samples, and to minimize the structural damage from the crystallization process. This may have been obvious from work in cryo-TEM (Adrian et al., 1984; Dubochet et al., 1988), but we hoped the damage would be limited to the ice structure and not that the ferritin particles would end up shredded and components dispersed. In the future, we expect to be able to perform vitrification of the ferritin-containing solution, through the use of a HPF or a plunge freezer into liquid ethane, and perform these experiments once again, in search for undamaged ferritin particles. Based on recent theoretical studies of the field evaporation behavior of water from crystalline and amorphous ice, significant differences should not be expected in the fragmentation and evaporation behavior (Segreto et al., 2022). Yet, despite the proximity to the metallic substrate, which should lead to relatively higher electrostatic field conditions, as discussed above, numerous larger protonated water clusters are detected. This will once again require some dedicated and systematic studies.

### 4.6   Perspective on our preliminary APT results

Even if this series of specimens were prepared from the same solution, following the same procedures (Eric V Woods et al., 2023), the distribution of the organic fragments (C-based) and other elements of interest varied from dataset to dataset. Our observations suggest that the protein shells of the ferritin particle were destroyed, mainly through the freezing process, even if we cannot ignore the possibility that damage from the FIB could have also contributed.

Let us now consider some of the critical aspects for future efforts in this direction to analyze hydrated frozen biological molecules.

First and foremost, the metal substrate should not react with the liquid/solution sample, which will help eliminate spurious peaks in the mass spectrum. This creates numerous spurious peaks that can make it more difficult to identify the organic fragments from the biomolecules themselves. These metallic species may also affect the biomolecules within the solution, change their conformation for instance, destroy or bind to them as discussed in Woods et al. (Woods et al., 2025). Interestingly, elements from the ferritin shell and core were not found within the pores of the nano-porous substrate, but in the solution standing above, and in a relatively higher concentration closer to the brass substrate, including in the form of metal-organic complexes.

Second, the species distribution observed in the reconstructed APT maps may be affected by the specimen preparation itself. During the FIB milling process, the ferritin distribution is not visible from SEM, and the energetic Ga-ions may cause damage to the ferritin structure and change its distribution. The initial selection of the lamella or when to stop the annular milling process are not guided by visual clues of the presence of the particles. As a result, the APT specimen may not contain any ferritin molecules. The presence of the detected Fe, Na, Cl and organic fragments (C-based) hints as to the presence of the ferritin in some of the datasets as least (Fig. 5). The damage from the Ga implantation is of interest to the biologist. While the damage layer has been estimated using cryo-TEM (Lucas & Grigorieff, 2023; Yang et al., 2023), obtaining the distribution of Ga in 3D would help better understand the mechanism of Ga implantation in frozen solutions/ice and damage at the nanoscale, and possibly help guide optimization processes for better insights into biological samples.

Third, several peaks could not be readily identified. There are numerous combinations of C, H, O and possibly N and P, that can fit with a given mass-to-charge-ratio. Conducting comparative studies of saline solutions with the same concentration as the ferritin solution, and dried ferritin samples could also provide valuable insights to refine the identification of the peaks. The implementation of machine-learning based methods for mass spectrum analysis interpretation could effectively help address the challenges of peak identification and overlap (Wei et al., 2021; Mikhalychev et al., 2020; Haley et al., 2015), but progress needs to be made in the theory of field evaporation, formation and dissociation of molecular ions and particularly from solvated molecules. Precious but limited insights have been gained from quantum chemistry and atomistic simulations (Schwarz et al., 2020, 2021; Wang et al., 2006; Dietrich et al., 2020; Segreto et al., 2022).

Fourth, a major challenge lies in identifying each peak unambiguously due to fact that each peak can correspond to molecules with the combinations of carbon, hydrogen, oxygen, and nitrogen (Perea et al., 2016), along with dissolved hydrated metal ions (Eric V Woods et al., 2023; Woods et al., 2025). Furthermore, in hydrated specimens, there can be overlaps between organic fragments and protonated water clusters. The interpretation of the mass spectra and reconstruction of each individual fragment into a biologically meaningful molecule will also require targeted studies in the future.

Finally, these aspects combine to make quantitative analysis, at the moment, beyond extremely challenging. It could have been possible to maybe use the number of Fe in the ferritin cores to try to determine the expected number of particles within the analyzed volume. However, this is not necessarily a constant and can depend on the sample itself (Melino et al., 1978; Niitsu et al., 1985). The commercial ferritin we used did not have specified what is expected. The difficulties in identifying each and every cluster ion, as indicated in Figure 4, also makes a relative assessment of the ratio of Fe to C inaccessible at this stage.

## 5   Conclusion

Although ferritin particles appeared to be a good candidate for exploring the hydrated biological APT specimen preparation method and optimizing the APT parameters due to its Fe core, which should provide a unique marker, several key aspects require attention to further advance the study of hydrated biological specimens using APT. Moderate pulse repetition rates (100–120 kHz) and laser pulse energies in the range of 60–80 pJ help minimize the background. The absolute value of the laser pulse energy depends on the specimen shape and size, and should hence be adjusted individually for each experiment. Nanoporous brass was shown to be a suitable substrate by withstanding up to high voltages with limited micro-fractures, yet it produces numerous metal-complexes peaks in the mass spectrum that could obscure the signal of interest. Fe, Na, Cl and C-containing fragments were consistently detected, in a relatively higher concentration closer to the brass substrate, including in the form of metal-complexes. Nonetheless, interpreting peaks remains challenging, particularly due to the combinations and overlaps of organic fragments, and this will require systematic studies on simpler systems to facilitate identification. We have demonstrated in principle that the workflow used for preparing specimens at cryogenic temperature can be used for samples prepared by HPF, which should, in the future allow for imaging undamaged and hydrated ferritin particles or other biologically-relevant macromolecules.

## Acknowledgements

SZ, EVW, TMS and BG are grateful for funding from the German Research Foundation (DFG) through the award of the Leibniz Prize 2020 to BG. TMS gratefully acknowledges the financial support of the Walter Benjamin Program of the DFG (Project No. 551061178). We are grateful for help from Sebastian Tacke and Stefan Raunser from the MPI for Molecular Physiology for the provision of the HPF samples. The authors thank Uwe Tezins, Christian Bross, and Andreas Sturm for their support at the FIB and APT facilities at MPIE. The two reviewers are deeply thanked for their rather extensive yet constructive comments, and careful proof reading.

# Atom probe tomography of hydrated biomacromolecules: preliminary results

Shuo Zhang[1], Leonardo Shoji Aota[1], Mahander P. Singh[1], Eric V. Woods[1], Fantine Périer Jouet[1], Tim M. Schwarz[1], Baptiste Gault[1,2,*].

[1] Max-Planck Institute for Sustainable Materials (formerly Eisenforschung), Max-Planck-Strasse 1, 40237 Düsseldorf, Germany

[2] Department of Materials, Royal School of Mines, Imperial College

[*] Corresponding author: b.gault@mpie.de

## Supplementary Information

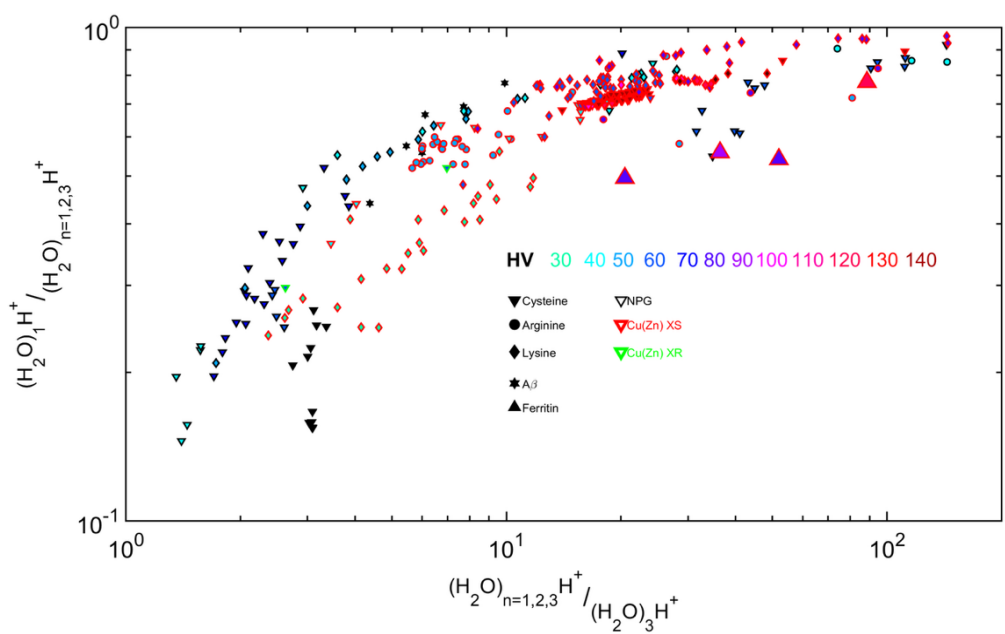

*Suppl. Figure 1: Graph of the CIR from Ref. (Woods et al., 2025) with addition of the data from the different sections of Figure 2b for the solution containing ferritin displayed with larger symbols.*

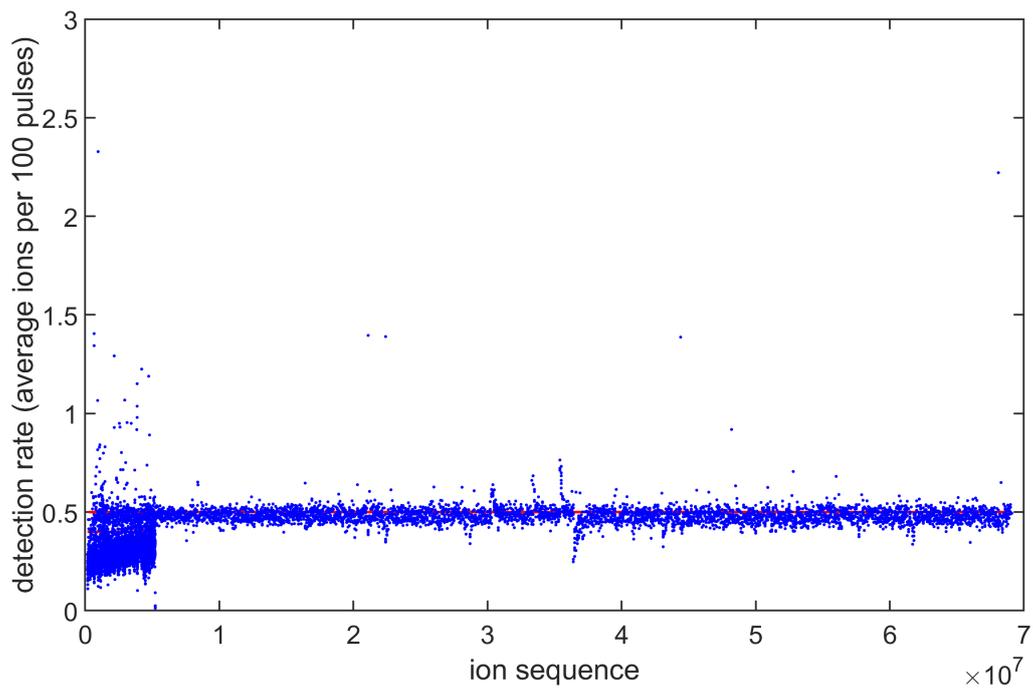

*Suppl. Figure 2: detection rate vs. ion sequence for the dataset in Figure 2b.*

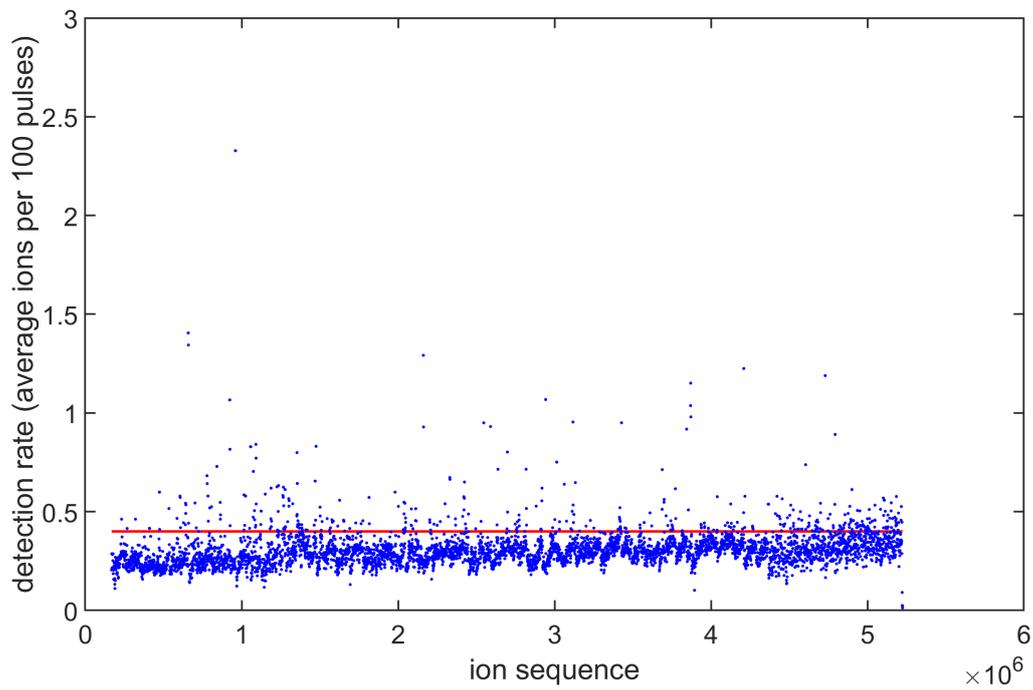

*Suppl. Figure 3: detection rate vs. ion sequence for the dataset in Figure 2c.*

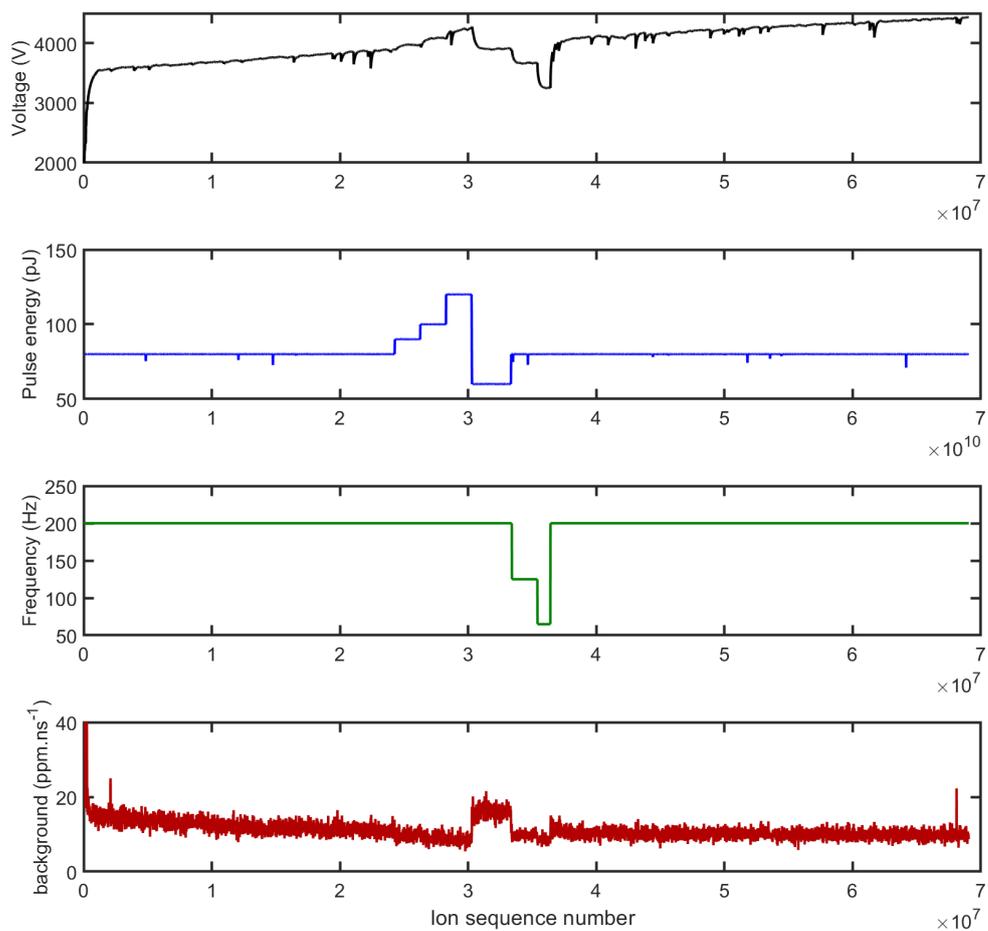

*Suppl. Figure 4: Evolution of the applied voltage, laser pulse energy, laser pulsing frequency and background for the dataset in Figure 2.*

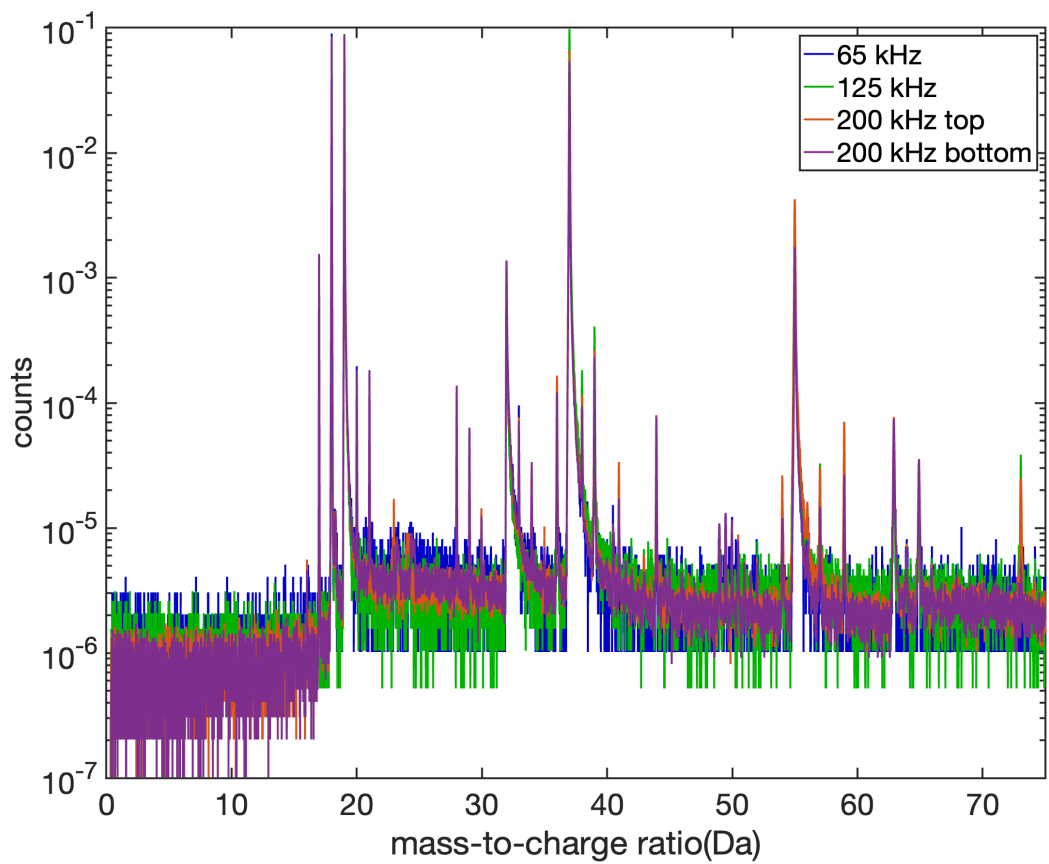

*Suppl. Figure 5: normalised mass spectra for the different regions of the dataset in Figure 2. Note that for these mass spectra a non-linear binning was used (the bin size increases quadratically from 0.001 to 0.04. This allows to have a background shaped similarly to that obtained on a time-of-flight spectrum.*

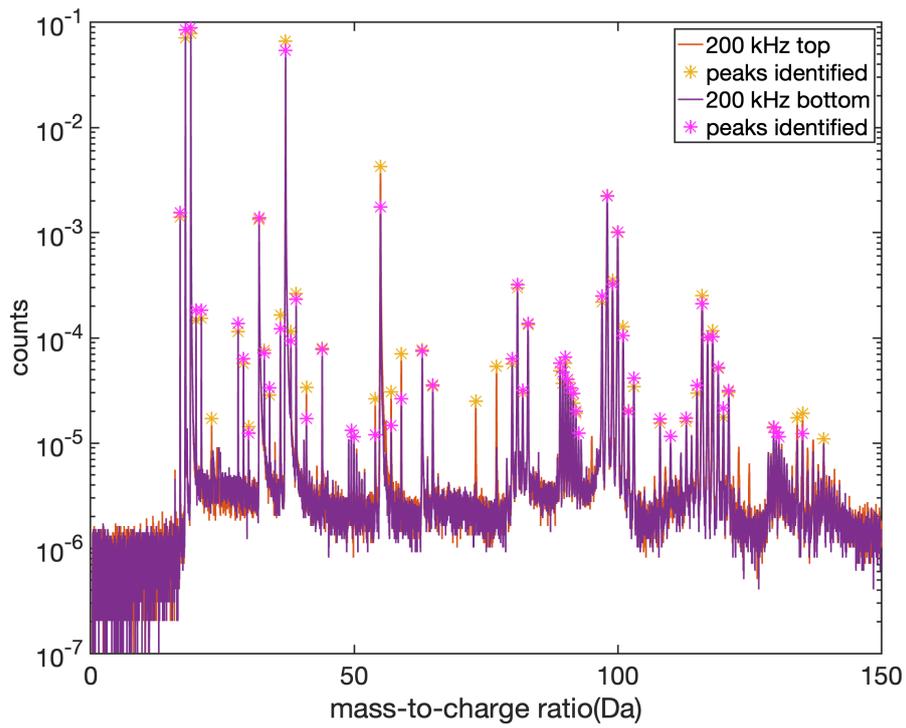

*Suppl. Figure 6: (a) mass spectra and peaks identified for the two spectra already showed in Figure 4a-b.*

| Amino acid | Symbol | Mass (Da) | Formula | Skeletal formula |
|---|---|---|---|---|
| **Glycine** | G | 75.1 | $C_2H_5NO_2$ | 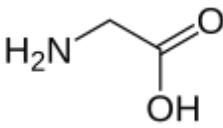 |
| **Alanine** | A | 89.1 | $C_3H_7NO_2$ | 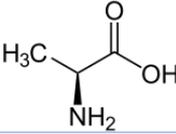 |
| **Serine** | S | 105.09 | $C_3H_7NO_3$ | 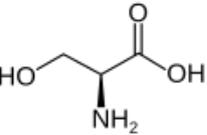 |
| **Valine** | V | 117.151 | $C_5H_{11}NO_2$ | 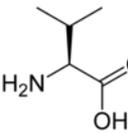 |
| **Threonine** | T | 119.11 | $C_4H_9NO_3$ | 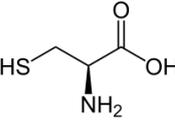 |
| **Cysteine** | C | 121.16 | $C_3H_7NO_2S$ | 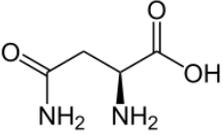 |
| **Asparagine** | Asn or N | 132.12 | $C_4H_8N_2O_3$ | 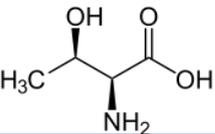 |
| **Aspartic acid** | Asp or D | 133.1 | $C_4H_7NO_4$ | 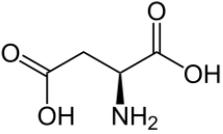 |
| **Glutamine** | Q | 146.146 | $C_5H_{10}N_2O_3$ | 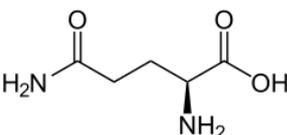 |
| **Histidine** | His or H | 155.1546 | $C_6H_9N_3O_2$ | 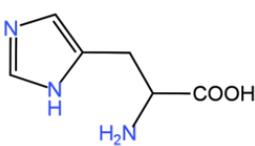 |
| **Tyrosine** | Y | 181.19 | $C_9H_{11}NO_3$ | 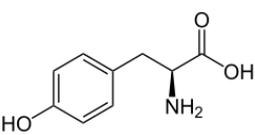 |

*Suppl. Table 1: selected amino acids with their mass, formula and skeletal formula.*

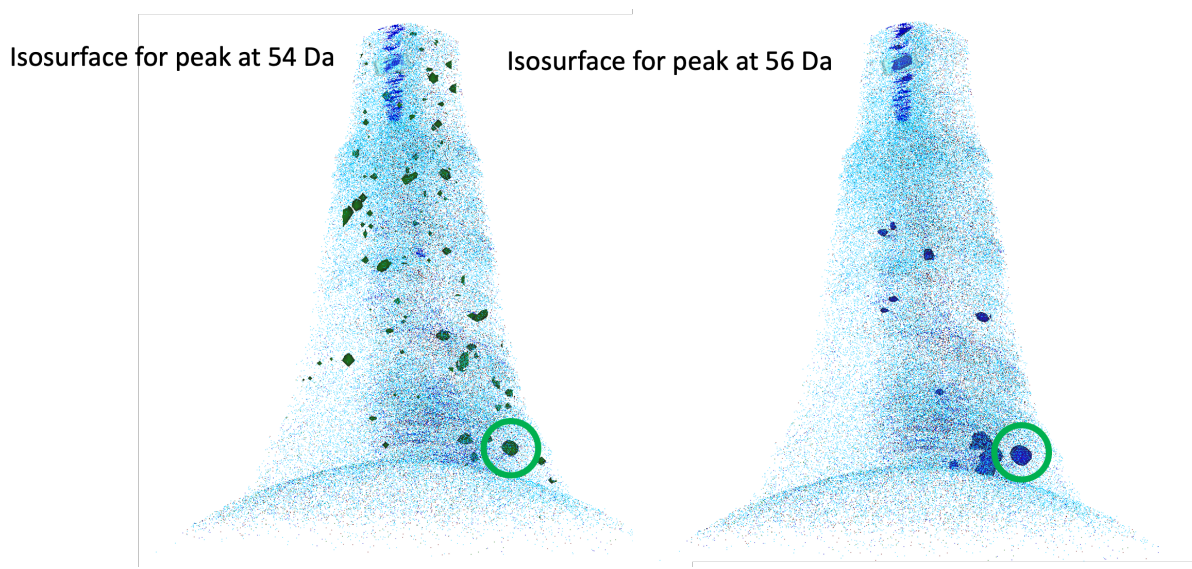

*Suppl. Figure 7: isosurfaces for the peak at 54 and 56 Da superimposed to the distribution of water cluster ions in the dataset shown in Figure 5.*

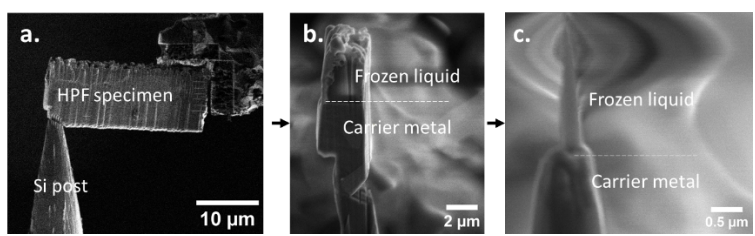

*Suppl. Figure 8: a) The lamella lifted out from the HPFed sample and b) attached to the Si supporting post and c) annular milled to obtain the final APT specimen.*

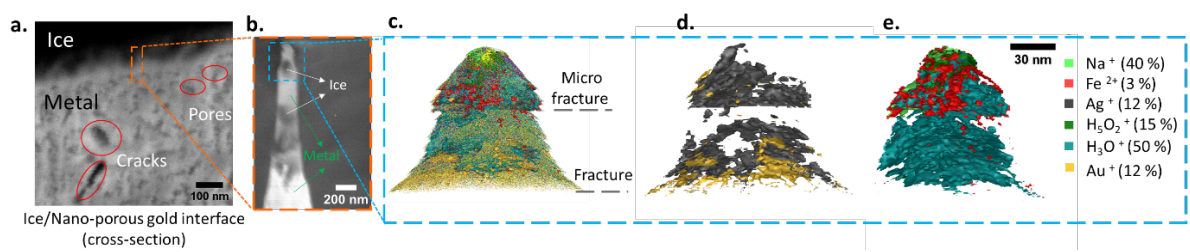

*Suppl. Figure 9: Cross-section shows the ice/NPG interface, and the porous structure of NPG. b) The APT ice specimen consists of ice and the underlying meta substratel, with a tip diameter ~100 nm. C) The isoconcentration surface of Na, Fe, C, Ag, and Au overlaps with the $H_3O$. The isoconcentration surface of d) the substrate metal and e) the elements from the frozen solution.*

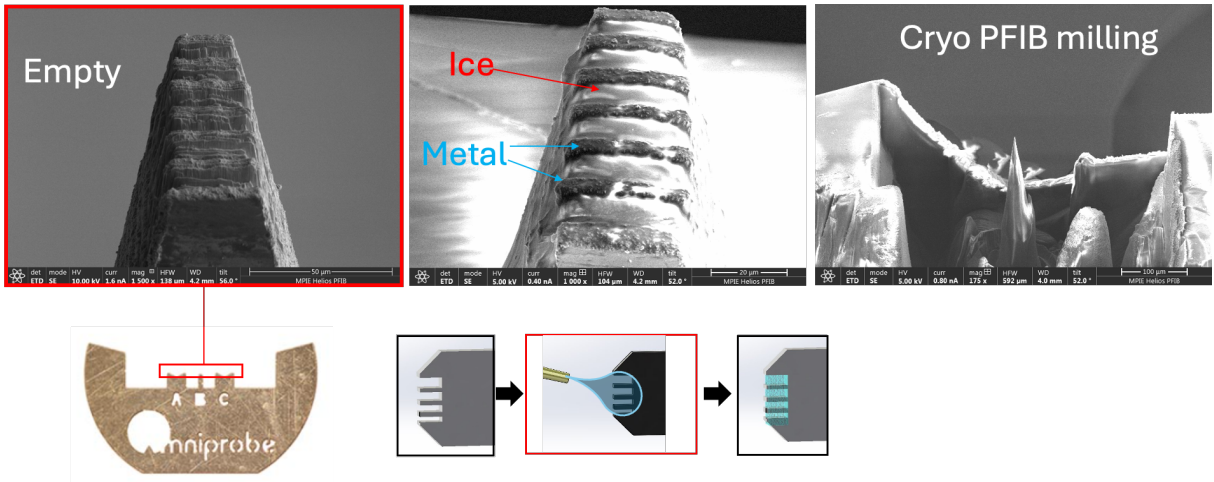

*Suppl. Figure 10: TEM grid observed in the SEM prior to wetting; grid carrying an aqueous solution after plunge freezing and imaged in the cryo-SEM/PFIB; scanning electron micrograph after one arm was shaped into a specimen for APT.*